\documentclass[preprint,showpacs,preprintnumbers]{revtex4}

\newcommand{\invfb}{\ensuremath{\text{fb}^{\text{-}1}}}

\usepackage{graphicx}
\usepackage{dcolumn}
\usepackage{bm}
\usepackage{epsfig}
\usepackage{epstopdf}
\usepackage{grffile}
\usepackage{color}
\usepackage{colordvi}
\usepackage{amssymb}
\usepackage{rotating}
\usepackage{lscape}
\usepackage{float}

\usepackage{footnote}
 \makesavenoteenv{environmentname}

\usepackage{hyperref}

\usepackage[T1]{fontenc} 

\begin{document}


\title{The LHC Drell-Yan Measurements as a Constraint for the Noncommutative Space-Time}

\author{Zahra Rezaei}
\email{zahra.rezaei@yazd.ac.ir}
\affiliation{Faculty of Physics, Yazd University, P.O. Box 89195-741, Yazd, Iran}

\author{Saeid Paktinat Mehdiabadi}
\email{paktinat@ipm.ir}
\affiliation{School of Particles and Accelerators, Institute for Research in Fundamental Sciences (IPM), P.O.Box 19395-5531, Tehran, Iran\\
	Faculty of Physics, Yazd University, P.O. Box 89195-741, Yazd, Iran}

\date{\today}

\begin{abstract}
The LHC measurements of the differential cross section of the Drell-Yan process is used to constrain the noncommutative space-time (NCST) phase space.
A $\chi^2$ method is utilized to exclude the part of the parameters space which is not consistent with the LHC measurements. Depending on other NCST parameters,
the scale can be pushed up to $\Lambda_{NC} > 655$ GeV. The effect of other parameters is also investigated. 
To our knowledge, it is the first time that a detailed statistical analysis is performed to compare the NCST results with the Hadron Collider measurements.
\end{abstract}
\maketitle

\section{Introduction}\label{sec:int} 
After the success of LHC to discover the last particle of the Standard Model (SM) puzzle, looking for possible signature of new physics beyond SM using the exact measurement of physical phenomena is put on the program. However no signal from new physics has been observed, so strong constraints are introduced by the LHC data \cite{Sirunyan:2017elk, Sirunyan:2018exx, Khachatryan:2016trj, Aaboud:2018zeb, Aaboud:2017gsl, Aaboud:2018urx}. The noncommutative (NC) SM is one of extensions of SM which considers the NC relation between the space-time coordinates and could be tested in LHC.  

However the preliminary idea of noncommutativity dates back to Heisenberg, with motivation coming from the string theory, the possibility of NC space-time (NCST) has received a great deal of attention. Since SM is supposed to be an effective low energy theory which has succeeded to describe the low energy experiments in the nature and the string theory is a considerable theory in the Planck scale, NCST, as a theory which contains both aspects, ought to be valid in intermediate energies.

The NCST leads to a commutation relation between space-time operators 
\begin{equation}
	[\hat{x}^{\mu},\hat{x}^{\nu}]=i\theta^{\mu\nu}=\frac{iC^{\mu\nu}}{\Lambda^2_{NC}},\label{eq1} 
\end{equation}
where the hatted quantities are the NC coordinates and $\theta^{\mu\nu}$ is a constant, real, antisymmetric tensor, called NC parameter, which has the dimension of $[mass]^{-2}$. In Eq. (\ref{eq1}), $\Lambda_{NC}$ is the energy scale of NCST and $C^{\mu\nu}$ is a dimensionless electromagnetic field strength like tensor. A simple way to NC field theory is the Weyl-Moyal star product \cite{Madore:2000en,Riad:2000vy}:
\begin{equation}
	(f * g)(x) = {\left. {\exp (\frac{1}{2}i{\theta ^{\mu \nu }}\,\frac{\partial }{{\partial {x^\mu }}}\frac{\partial }{{\partial {y^\nu }}})f(x)\,g(y)} \right|_{y \to x}}
	\label{equation.2}
\end{equation}
Substituting star product for usual multiplication between conventional fields will lead to NC field theory. This mechanism makes some difficulties such as charge quantization \cite{Hayakawa:1999yt,Hayakawa:1999zf} and definition of gauge group tensor product \cite{Chaichian:2001mu}. To solve these problems, two approaches are assumed. The first one is built from a $U(n)$ gauge group which is larger than the SM one and it reduces to the SM gauge group using two Higgs mechanism \cite{Chaichian:2001py}. The Seiberg-Witten map \cite{Seiberg:1999vs} is the second approach which has the gauge group like the SM one and the NC fields are expanded in terms of the commutative fields \cite{Calmet:2001na}. 

In both approaches, existing vertices receive corrections in the leading and higher order of NC parameter $\theta^{\mu\nu}$. Moreover, usually SM forbidden interactions such as triple neutral gauge boson couplings ($\gamma gg$, Z$\gamma\gamma$, Z$gg$ and $\gamma\gamma\gamma$) are allowed in NCST. The range of all coupling constant values, has been completely computed in \cite{Behr:2002wx,Duplancic:2003hg}. 

In a substantial number of articles, the phenomenological aspects of NCST have been studied to find possible experimental indications and/or estimate bounds on the theory from the experimental data in both approaches. These experiments include the low energy as well as the high energy collider experiments and the astrophysical events. 

At high energy collider experiments, different processes  have been investigated in high energy $e^- e^+$ colliders as well as hadron colliders. The $e^- e^+$ scattering in different channels (Bhabha, Moller and Compton scattering, pair annihilation and pair production) 
have been investigated. Study on $e^- e^+$ processes in future colliders demonstrates that the reachable scale of NC is proportional to the collider's center-of-mass energy ($\Lambda_{NC}\sim \sqrt{s}$) \cite{Hewett:2002zp,Mathews:2000we,Baek:2001ty} and significant deviation of physical observables from the SM ones has been estimated for a few TeV. The OPAL collaboration has already examined the effect of NCST on pair annihilation with the LEP data \cite{Abbiendi:2003wva}. Using distributions of the photon angle, a lower limit on the energy scale $\Lambda_{NC}$ of $141~$GeV at the $95\%$ confidence level is extracted which is much smaller than the scale of the future high energy $e^- e^+$ colliders. This is the only analysis that uses the experimental data from the colliders to constrain the NC scale.

The production cross section of the t-channel single top quark at LHC is studied in  NCST \cite{YaserAyazi:2012ni}. The authors constrain $\Lambda_{NC}$ by using the predicted experimental precision on spin correlation. Studying the production of neutral vector bosons Z$\gamma$ at LHC, under some conservative assumptions, has suggested the physical scale $\Lambda_{NC}\simeq 1~$TeV to observe the effect of NC nature of the space-time \cite{Alboteanu:2006hh}. Also according to pair production of charged gauge bosons at LHC, it is predicted that the azimuthal oscillation is visible for $\Lambda_{NC}=700~$GeV \cite{Ohl:2010zf}. 

The di-lepton final state of the Drell-Yan (DY) processes from $pp/p\bar{p}$ collisions are appropriate channels to test new physics in the light of clearness of these channels. We study the NCST effect on the cross section of di-leptons to constrain the NC parameters space using the LHC data. This process has been investigated in NCST in Ref \cite{Selvaganapathy:2016jrl}. 

The effect of NCST on the proton structure functions at electron-proton colliders has also been studied \cite{Rafiei:2016lqq}. The obtained results for the improved proton structure function are in better compatibility with the available experimental data rather than the results coming from the normal parametrization models, especially at the high energy region, for the NC scale $\Lambda_{NC}\gtrsim 0.4~$TeV. 


In this paper, we concentrate on the DY processes in the LHC experiments to determine the allowed parameters space of NCST. To accomplish it, we will utilize the LHC data in $\sqrt{s}=7,~8$ TeV. The structure of the paper is as follows: In Sec.~\ref{sec:ncst} we discuss the DY processes in NCST. The experimental data is presented in Sec.~\ref{sec:exp}. In Sec.~\ref{sec:res} the method and results of the analysis are presented. Finally, Sec.~\ref{sec:con} summarizes the paper.

\section{Drell-Yan processes in NCST}\label{sec:ncst} 
The DY processes involve the production of lepton pairs which have been measured precisely in hadronic collisions. Duo to large event rates and its clean final state, which stands out from colour charges, the DY processes are of great importance to search for new physics beyond SM. Measurements of the production rates and masses of W and Z bosons, test of parton model ideas, determining the parton distributions inside a nucleon, discovering or limiting new physics involving heavier particles similar to the W and Z bosons are part of DY applications through the recent five decades. 

In SM, lepton pairs are produced from quark-antiquark pair annihilation of initial protons mediated by vector gauge bosons:
\begin{equation} q\bar{q}\longrightarrow \gamma^*/ Z \longrightarrow l^- l^+ .
\end{equation}
According to the triple neutral gauge boson couplings in NCST, there are additional channels to produce the final state lepton pairs in this theory as:
\begin{equation} gg\longrightarrow \gamma^*/ Z \longrightarrow l^- l^+.\nonumber
\end{equation}
The photon-gluon and Z-gluon couplings are indicated by $K_{\gamma gg}$ and $K_{Zgg}$ respectively, and their definitions have been shown in Ref \cite{Behr:2002wx,Duplancic:2003hg}. The couplings are not independent, $K_{Zgg}  = -\tan \theta_{w} K_{\gamma gg}$, where $\theta_{w}$ is the Weinberg angle, and below only the variation of $K_{Zgg}$ is considered.

The cross section of the DY process is calculated by convolution of the modified vertices and the newly introduced ones by NCST in Ref. \cite{Selvaganapathy:2016jrl}. They have depicted the differential cross section versus the invariant  mass, $\Lambda_{NC}$ and $\sqrt{s}$. 
Calculated amplitude and cross section are proportional to the NC parameter $\theta_{\mu\nu}$. Since  $\theta_{\mu 0} \neq 0 $ does not preserve the unitarity of S matrix, we choose $\theta_{\mu 0} = 0 $. Therefore we define $\theta_{ij}=\vec{\theta}=(\theta_{23},\theta_{31},\theta_{12})$ considering $\theta'$ and $ \phi'$ as the polar and azimuthal angles of $\vec{\theta}$. To include the contribution of $\theta_{\mu 0} \neq 0 $, one ought to consider the effect of earth rotation.  

\section{Experimental Data}\label{sec:exp} 
Both CMS and ATLAS experiments at LHC have measured the cross section of DY process in different $\sqrt{s}$. The measurements include both the inclusive and differential cross sections. In some cases, the measurements are done in double or even triple differential basis. In this analysis, due to the use of a leading order (LO) calculation, the measurements, which provide the cross section in the sub ranges of the mediator mass, are used. The CMS experiment has used more than 4.5 \invfb of data in the $\sqrt{s}$ = 7 TeV to measure the differential ($\frac{d\sigma}{dm}$) and double  differential ($\frac{d^2\sigma}{dmdy}$) cross section, where $m$ and $y$ are the mediator mass and rapidity, respectively \cite{Chatrchyan:2013tia}. The differential cross section measurement covers the mass range 15-1500 GeV.  Similar results are also provided by using 19.7 \invfb of data in the $\sqrt{s}$ = 8 TeV \cite{CMS:2014jea}. This measurement has reported the differential cross section up to 2000 GeV. 

The ATLAS experiment, has reported the differential cross section of DY process in  $\sqrt{s}$ = 8 TeV for low mass \cite{Aaboud:2017ffb}  and high mass \cite{Aad:2016zzw} dileptons, separately. The former analysis uses 20.2 \invfb and covers the mass range of 46-200 GeV. The mass range 116-1500 GeV is covered by the latter analysis which uses 20.3 \invfb of data. 

Apart from the 7 TeV analysis of the CMS experiment which uses the mass range 60-120 GeV, as the Z peak and reports the cross section normalized to the Z peak region, other measurements are the absolute values, without normalization. For the ATLAS result, due to the different binning that they have used, the mass range, 66-116 GeV is used as the Z peak region. The results are reported for dimuon and dielectron channels and also their combination. This analysis uses the latter results. The measurements are reported after correcting to the full phase space and easily can be compared to the output of our calculation without any need to consider the detector effects.

\section{Analysis and Results}\label{sec:res} 
\begin{figure}[!htb]
	\centering
	\includegraphics*[width=.45\textwidth]{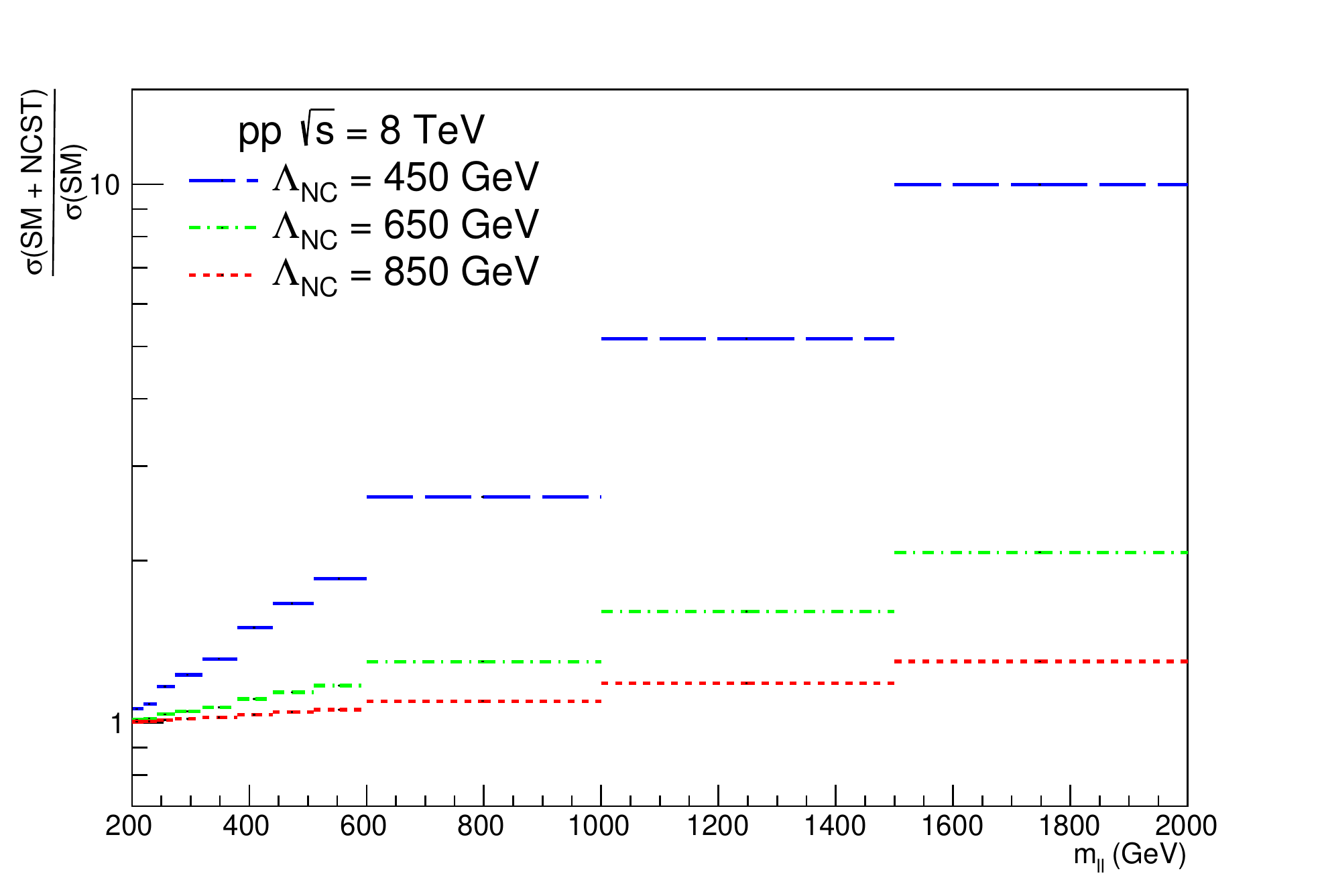}
	\caption{The ratio of the cross sections when NCST is considered and when it is ignored is shown for different scales of NC. The calculations are done for $\theta '=\phi '=\frac{\pi}{2}$  and $K_{Zgg} = 0.217$.}
	\label{fig:NCSTOverSM_dmll_8TeV}
\end{figure}

\begin{figure}[!htb]
	\centering
	\includegraphics*[width=.45\textwidth]{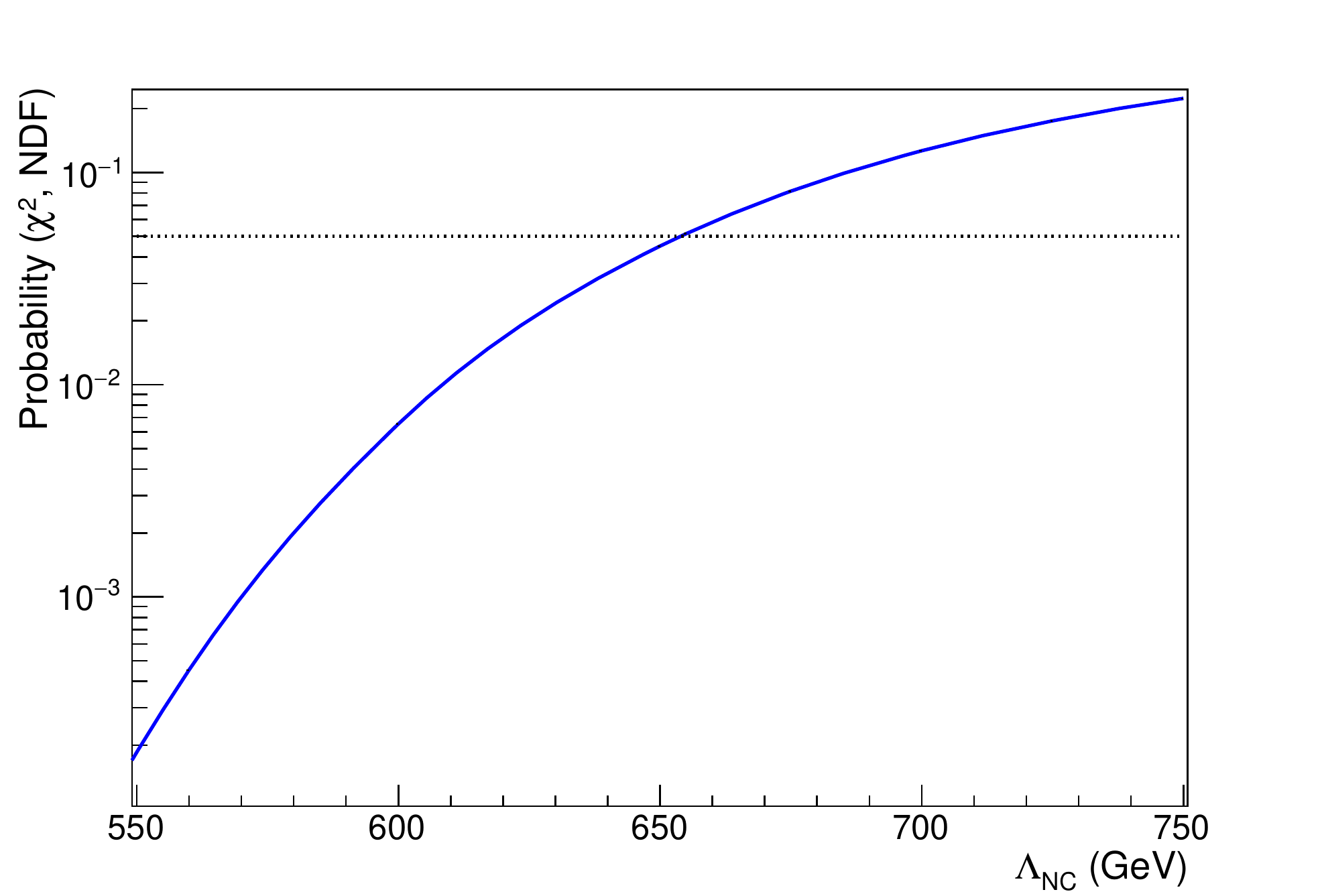}
	\caption{The $\chi^2$ probability versus $\Lambda_{NC}$. Other parameters are similar to Fig. \ref{fig:NCSTOverSM_dmll_8TeV}. The scales below 655 GeV are excluded at 95\% confidence level.}
	\label{fig:ExclusionvsLambdaNC}
\end{figure}

The matrix elements and spin-averaged squared amplitudes  of DY process when the NCST interactions are also considered, can be found in Ref. \cite{Selvaganapathy:2016jrl}. A C++ program is written to implement this squared amplitude and integrate over the phase space and the proton parton distribution functions (PDF). The CTEQ5L PDF is employed throughout the analysis when the $Q^2$ is set to the square of the Z boson mass ($Q^2$ = $91^2$ GeV$^2$). The matrix element is in LO of QCD calculations so the cross section does not include the higher order corrections and difference with the measured cross section is expected. Fortunately, 
the shape of the differential cross section as a function of the dilepton invariant mass is consistent with the measured spectrum in the experiments. For the normalization, the Z peak region is used and the calculated cross section is scaled to have the same integral in the Z peak region in both calculated and measured cross sections. A $\chi^2$ is defined to test the consistency of two distributions.
\begin{equation}
	\chi^2 = \sum_i \frac{(\sigma_{meas}^i - \sigma_{calc}^i)^2}{\delta_{i, meas}^2 + \delta_{i, calc}^2}
	\label{eq:chi2}
\end{equation}
The sum goes over the bins which are under the test of consistency and the number of the bins defines the number of degrees of freedom (NDF). The measured and calculated cross sections are shown as $\sigma_{meas}$ and $\sigma_{calc}$ and their uncertainties are shown with $\delta$. The measured uncertainties ($\delta_{meas}$) are taken from the experimental paper and the calculated uncertainty is taken as 20\% of the calculated cross section. This pessimistic value should cover different theoretical uncertainties like PDF, higher order corrections and $Q^2$ scale. In Fig. \ref{fig:NCSTOverSM_dmll_8TeV} 
the ratio of the cross sections when NCST is on and when only SM is considered are plotted versus the dilepton invariant mass for $\sqrt{s}=8$ TeV. It can be seen that the deviation from 1 starts to get visible, above $m_{ll}$ = 200 GeV. Hence the defined $\chi^2$ is used for the bins above 200 GeV for 3 different measurements which are listed in Sec. \ref{sec:exp} (CMS 7 and 8 TeV, ATLAS 8 TeV). Some deviation also exist in the Z peak region, but since we use this region for normalization, it can not be used as the signal region. The number of bins which sets NDF, is 29. When the pure SM is considered, the $\chi^2$ is 26.53 and its probability is above 60\%. It confirms that the calculated cross section is consistent with the measured cross section within the uncertainties. To find the part of the NCST phase space which is inconsistent with data, the NCST parameters are changed and the cross section is calculated. Then $\chi^2$ is examined, if the $\chi^2$ probability is less than 5\%, that part of the phase space is considered to be inconsistent with data and it is excluded at 95\% confidence level. The statistical method follows the prescription of Ref. \cite{Feldman:1997qc}. In Fig. \ref{fig:ExclusionvsLambdaNC}
the $\chi^2$ probability is shown when the $\Lambda_{NC}$ is changing. For  $\Lambda_{NC} <$ 655 GeV, the probability is less than 5\%, so we conclude that NCST scale must be greater than this value, if it is the true theory to explain the nature.

\begin{figure}[!htb]
	\centering
	\includegraphics*[width=.45\textwidth]{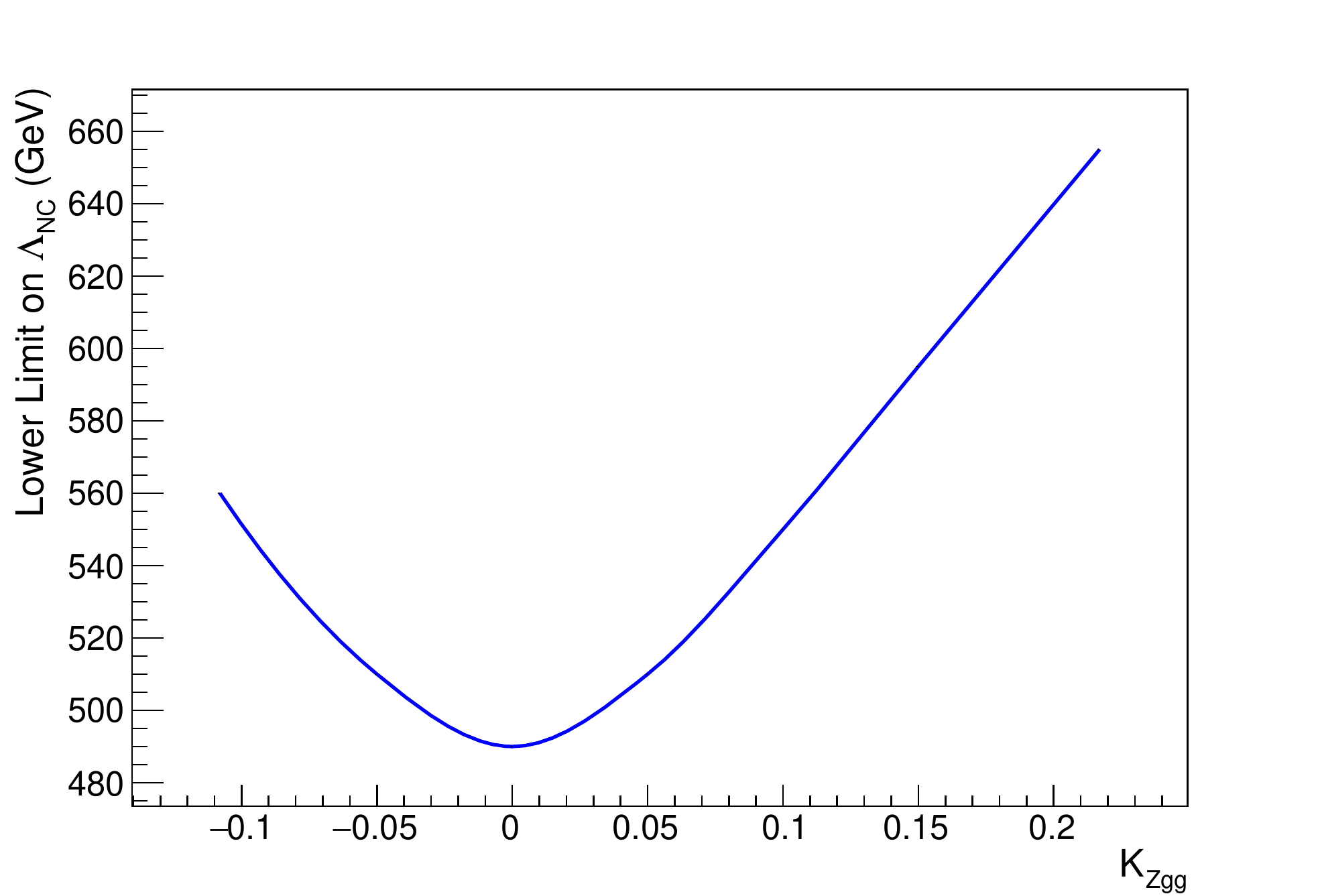}
	\includegraphics*[width=.45\textwidth]{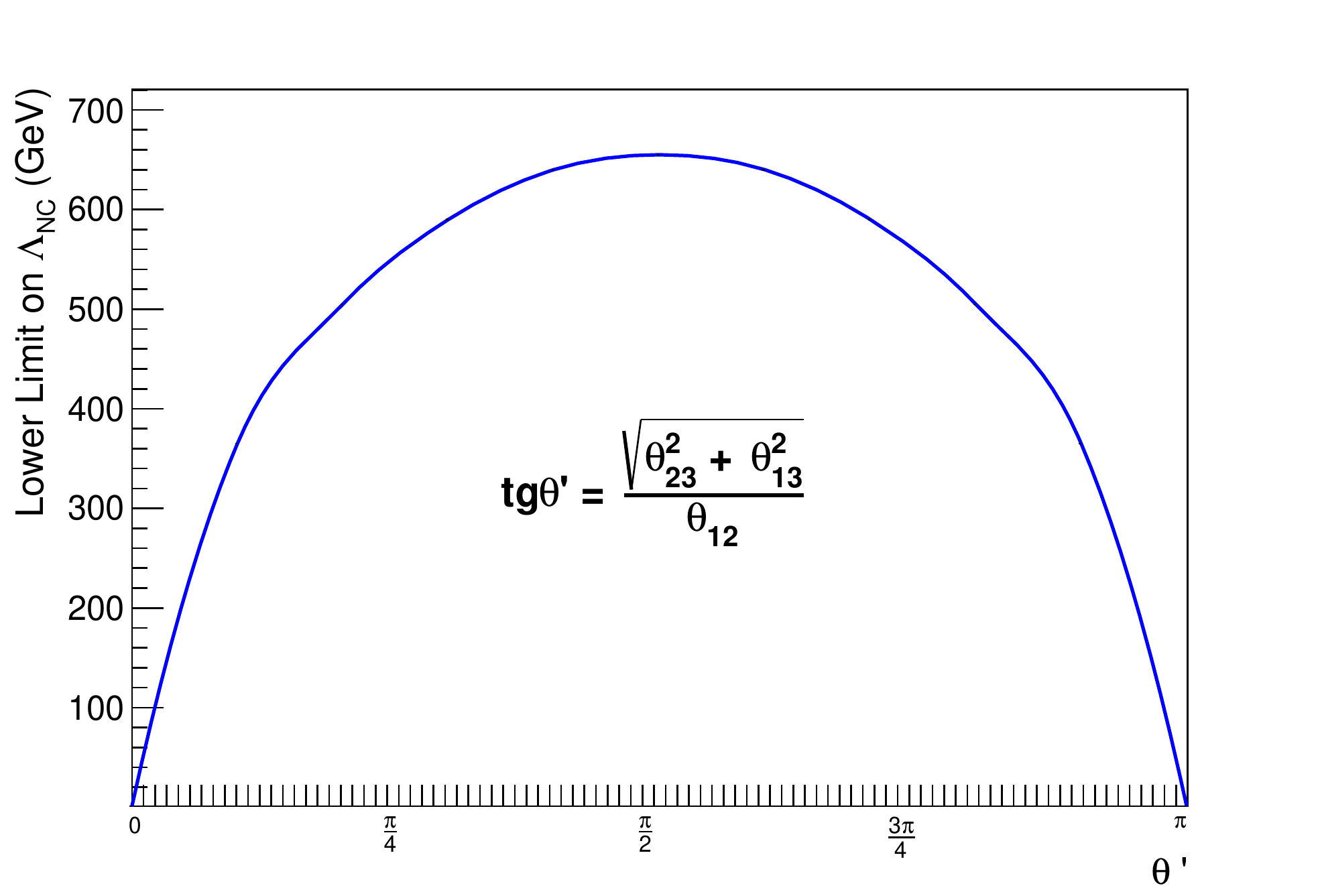}
	\caption{The lower limit on $\Lambda_{NC}$ when other parameters are changing. In left, $\theta '$ is fixed at $\frac{\pi}{2}$ and $K_{Zgg}$ is varying. In right, $K_{Zgg}$ is set equal to 0.217 and $\theta '$ is scanned.}
	\label{fig:ExclusionvsKzggThetaThetaNC}
\end{figure}

\begin{figure}[!htb]
	\centering
	\includegraphics*[width=.45\textwidth]{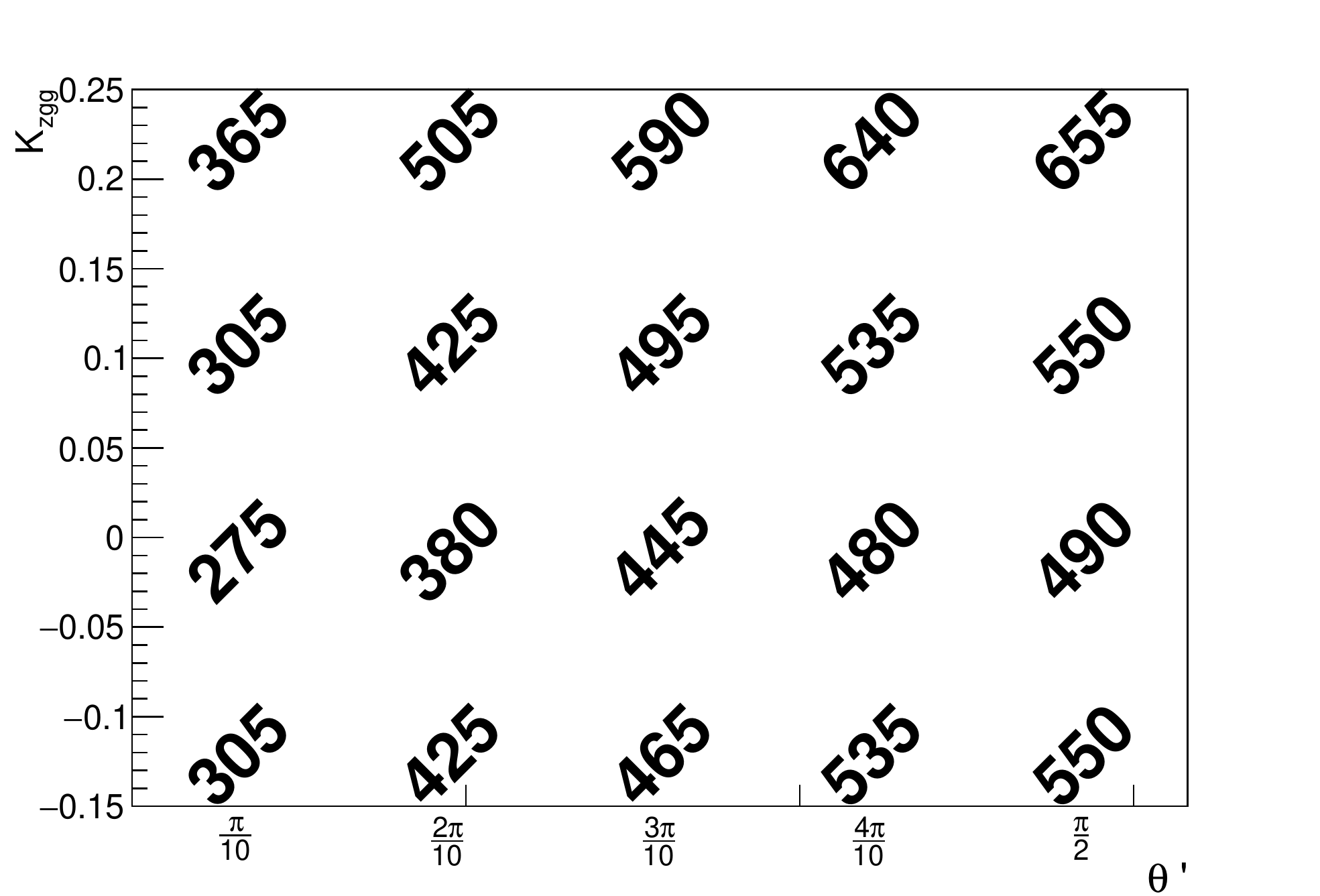}
	\caption{The lower limit on $\Lambda_{NC}$ when other parameters are changing simultaneously. No correlation is seen between $K_{Zgg}$ and $\theta '$.}
	\label{fig:ExclusionKzggThetaThetaNC}
\end{figure}

In next step, the effect of different parameters on the exclusion is investigated. In Fig. \ref{fig:ExclusionvsKzggThetaThetaNC}
the excluded NCST scale is plotted for the allowed region of $K_{Zgg}$ and $\theta '$. 
The new interactions introduced by NCST are proportional to $K_{Zgg}$ and cross section is related to the square of this value, so a parabolic curve is seen, when $K_{Zgg}$ is changing. Dependency of the scale lower limit on $\theta '$ is symmetric as it is expected for a totally symmetric collider like LHC which is a pp collider. When  $\theta '$ is close to 0 or $\pi$, the analysis can not exclude any scale, because the NCST contribution in total cross section is zero. The best limits are found when $K_{Zgg}$ is equal to 0.217 or $\theta '$ is equal to $\frac{\pi}{2}$. To investigate any correlation between these two variables, the lower limit is found when they are changing simultaneously. The result is shown in Fig.  \ref{fig:ExclusionKzggThetaThetaNC}.
No correlation is seen between $K_{Zgg}$ and $\theta '$, as expected. The effect of changing $\phi '$ on the lower limit is also studied, but no dependency is seen on this parameter. For all plots, the value of $\phi '$ is set equal to $\frac{\pi}{2}$, meaning that $\vec{\theta}$ is equal to (0, 1, 0).

\begin{figure}[!htb]
	\centering
	\includegraphics*[width=.45\textwidth]{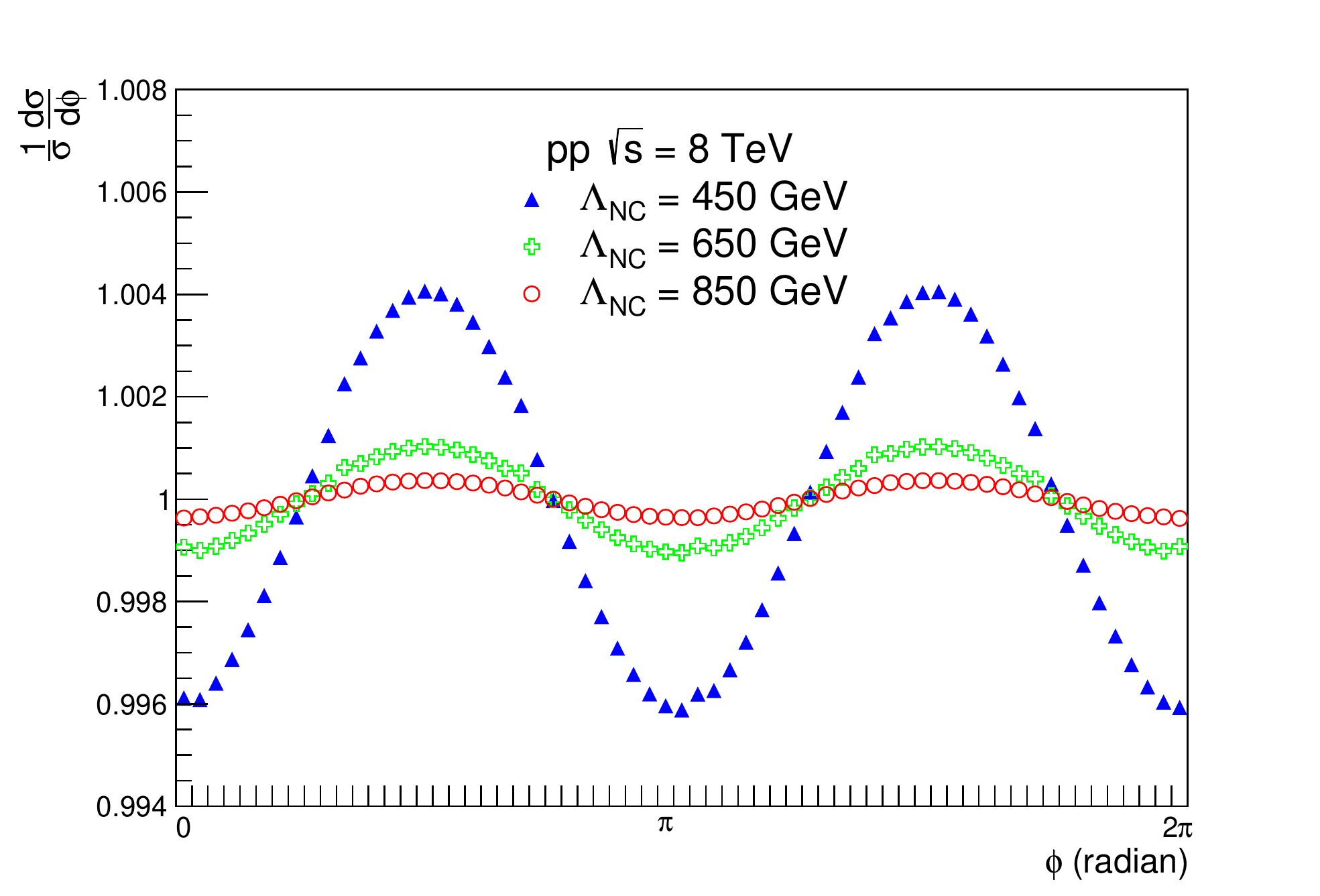}
	\caption{The normalized cross section versus the $\phi$ of the final lepton for the events under the Z peak. The oscillating azimuthal distribution is a unique feature of NCST.}
	\label{fig:NCSTOverSM_dphi_8TeV}
\end{figure}

The main feature of NCST is the violation of the Lorentz invariance, that can be seen especially, in the azimuthal distribution around the beam direction. Figure \ref{fig:NCSTOverSM_dphi_8TeV}
shows the normalized cross section versus the $\phi$ of the final lepton for the events with $60 < m_{ll} < 120$ GeV. For the pure SM contribution in DY process, a horizontal line with intercept of 1 is expected. The experiments at LHC or future colliders can try to decrease the uncertainties on this distribution and look for any deviation from the horizontal line as a unique signature of new physics beyond SM.

\section{Conclusion}\label{sec:con} 
For the first time a detailed statistical analysis is performed to compare the NCST results with the Hadron Collider measurements. The NCST results are evaluated by a matrix element calculation implemented in a private code for integration on the phase space and PDF. The differential cross section as a function of the mediator invariant mass is used to look for any excess due to NCST effects. A $\chi^2$ measure is utilized to quantify the deviation from the LHC measurements. Depending on other NCST parameters, $\Lambda_{NC} < 655$ GeV is excluded at 95\% confidence level. The effect of other parameters on the exclusion is studied. The oscillation in the azimuthal angle of the final lepton is suggested as a unique signature of the new model that can be measured at LHC or the future colliders.

\section{Acknowledgments}
The authors would like to thank Seyed Yaser Ayazi, for the useful discussions and his comments on the manuscript. 
The authors are grateful to the CMS and ATLAS collaborations for their fantastic results.


\begin{thebibliography}{30}
	\expandafter\ifx\csname natexlab\endcsname\relax\def\natexlab#1{#1}\fi
	\expandafter\ifx\csname bibnamefont\endcsname\relax
	\def\bibnamefont#1{#1}\fi
	\expandafter\ifx\csname bibfnamefont\endcsname\relax
	\def\bibfnamefont#1{#1}\fi
	\expandafter\ifx\csname citenamefont\endcsname\relax
	\def\citenamefont#1{#1}\fi
	\expandafter\ifx\csname url\endcsname\relax
	\def\url#1{\texttt{#1}}\fi
	\expandafter\ifx\csname urlprefix\endcsname\relax\def\urlprefix{URL }\fi
	\providecommand{\bibinfo}[2]{#2}
	\providecommand{\eprint}[2][]{\url{#2}}
	
	\bibitem[{\citenamefont{Sirunyan
			et~al.}(2018{\natexlab{a}})}]{Sirunyan:2017elk}
	\bibinfo{author}{\bibfnamefont{A.~M.} \bibnamefont{Sirunyan}}
	\bibnamefont{et~al.} (\bibinfo{collaboration}{CMS}), \bibinfo{journal}{Phys.
		Lett.} \textbf{\bibinfo{volume}{B780}}, \bibinfo{pages}{501}
	(\bibinfo{year}{2018}{\natexlab{a}}), \eprint{1709.07497}.
	
	\bibitem[{\citenamefont{Sirunyan
			et~al.}(2018{\natexlab{b}})}]{Sirunyan:2018exx}
	\bibinfo{author}{\bibfnamefont{A.~M.} \bibnamefont{Sirunyan}}
	\bibnamefont{et~al.} (\bibinfo{collaboration}{CMS}), \bibinfo{journal}{JHEP}
	\textbf{\bibinfo{volume}{06}}, \bibinfo{pages}{120}
	(\bibinfo{year}{2018}{\natexlab{b}}), \eprint{1803.06292}.
	
	\bibitem[{\citenamefont{Khachatryan et~al.}(2017)}]{Khachatryan:2016trj}
	\bibinfo{author}{\bibfnamefont{V.}~\bibnamefont{Khachatryan}}
	\bibnamefont{et~al.} (\bibinfo{collaboration}{CMS}), \bibinfo{journal}{JHEP}
	\textbf{\bibinfo{volume}{04}}, \bibinfo{pages}{018} (\bibinfo{year}{2017}),
	\eprint{1610.04870}.
	
	\bibitem[{\citenamefont{Aaboud et~al.}(2018{\natexlab{a}})}]{Aaboud:2018zeb}
	\bibinfo{author}{\bibfnamefont{M.}~\bibnamefont{Aaboud}} \bibnamefont{et~al.}
	(\bibinfo{collaboration}{ATLAS}), \bibinfo{journal}{Phys. Rev.}
	\textbf{\bibinfo{volume}{D98}}, \bibinfo{pages}{032009}
	(\bibinfo{year}{2018}{\natexlab{a}}), \eprint{1804.03602}.
	
	\bibitem[{\citenamefont{Aaboud et~al.}(2018{\natexlab{b}})}]{Aaboud:2017gsl}
	\bibinfo{author}{\bibfnamefont{M.}~\bibnamefont{Aaboud}} \bibnamefont{et~al.}
	(\bibinfo{collaboration}{ATLAS}), \bibinfo{journal}{Eur. Phys. J.}
	\textbf{\bibinfo{volume}{C78}}, \bibinfo{pages}{24}
	(\bibinfo{year}{2018}{\natexlab{b}}), \eprint{1710.01123}.
	
	\bibitem[{\citenamefont{Aaboud et~al.}(2018{\natexlab{c}})}]{Aaboud:2018urx}
	\bibinfo{author}{\bibfnamefont{M.}~\bibnamefont{Aaboud}} \bibnamefont{et~al.}
	(\bibinfo{collaboration}{ATLAS}), \bibinfo{journal}{Phys. Lett.}
	\textbf{\bibinfo{volume}{B784}}, \bibinfo{pages}{173}
	(\bibinfo{year}{2018}{\natexlab{c}}), \eprint{1806.00425}.
	
	\bibitem[{\citenamefont{Madore et~al.}(2000)\citenamefont{Madore, Schraml,
			Schupp, and Wess}}]{Madore:2000en}
	\bibinfo{author}{\bibfnamefont{J.}~\bibnamefont{Madore}},
	\bibinfo{author}{\bibfnamefont{S.}~\bibnamefont{Schraml}},
	\bibinfo{author}{\bibfnamefont{P.}~\bibnamefont{Schupp}}, \bibnamefont{and}
	\bibinfo{author}{\bibfnamefont{J.}~\bibnamefont{Wess}},
	\bibinfo{journal}{Eur. Phys. J.} \textbf{\bibinfo{volume}{C16}},
	\bibinfo{pages}{161} (\bibinfo{year}{2000}), \eprint{hep-th/0001203}.
	
	\bibitem[{\citenamefont{Riad and Sheikh-Jabbari}(2000)}]{Riad:2000vy}
	\bibinfo{author}{\bibfnamefont{I.}~\bibnamefont{Riad}} \bibnamefont{and}
	\bibinfo{author}{\bibfnamefont{M.~M.} \bibnamefont{Sheikh-Jabbari}},
	\bibinfo{journal}{JHEP} \textbf{\bibinfo{volume}{08}}, \bibinfo{pages}{045}
	(\bibinfo{year}{2000}), \eprint{hep-th/0008132}.
	
	\bibitem[{\citenamefont{Hayakawa}(2000)}]{Hayakawa:1999yt}
	\bibinfo{author}{\bibfnamefont{M.}~\bibnamefont{Hayakawa}},
	\bibinfo{journal}{Phys. Lett.} \textbf{\bibinfo{volume}{B478}},
	\bibinfo{pages}{394} (\bibinfo{year}{2000}), \eprint{hep-th/9912094}.
	
	\bibitem[{\citenamefont{Hayakawa}(1999)}]{Hayakawa:1999zf}
	\bibinfo{author}{\bibfnamefont{M.}~\bibnamefont{Hayakawa}}
	(\bibinfo{year}{1999}), \eprint{hep-th/9912167}.
	
	\bibitem[{\citenamefont{Chaichian et~al.}(2002)\citenamefont{Chaichian,
			Presnajder, Sheikh-Jabbari, and Tureanu}}]{Chaichian:2001mu}
	\bibinfo{author}{\bibfnamefont{M.}~\bibnamefont{Chaichian}},
	\bibinfo{author}{\bibfnamefont{P.}~\bibnamefont{Presnajder}},
	\bibinfo{author}{\bibfnamefont{M.~M.} \bibnamefont{Sheikh-Jabbari}},
	\bibnamefont{and} \bibinfo{author}{\bibfnamefont{A.}~\bibnamefont{Tureanu}},
	\bibinfo{journal}{Phys. Lett.} \textbf{\bibinfo{volume}{B526}},
	\bibinfo{pages}{132} (\bibinfo{year}{2002}), \eprint{hep-th/0107037}.
	
	\bibitem[{\citenamefont{Chaichian et~al.}(2003)\citenamefont{Chaichian,
			Presnajder, Sheikh-Jabbari, and Tureanu}}]{Chaichian:2001py}
	\bibinfo{author}{\bibfnamefont{M.}~\bibnamefont{Chaichian}},
	\bibinfo{author}{\bibfnamefont{P.}~\bibnamefont{Presnajder}},
	\bibinfo{author}{\bibfnamefont{M.~M.} \bibnamefont{Sheikh-Jabbari}},
	\bibnamefont{and} \bibinfo{author}{\bibfnamefont{A.}~\bibnamefont{Tureanu}},
	\bibinfo{journal}{Eur. Phys. J.} \textbf{\bibinfo{volume}{C29}},
	\bibinfo{pages}{413} (\bibinfo{year}{2003}), \eprint{hep-th/0107055}.
	
	\bibitem[{\citenamefont{Seiberg and Witten}(1999)}]{Seiberg:1999vs}
	\bibinfo{author}{\bibfnamefont{N.}~\bibnamefont{Seiberg}} \bibnamefont{and}
	\bibinfo{author}{\bibfnamefont{E.}~\bibnamefont{Witten}},
	\bibinfo{journal}{JHEP} \textbf{\bibinfo{volume}{09}}, \bibinfo{pages}{032}
	(\bibinfo{year}{1999}), \eprint{hep-th/9908142}.
	
	\bibitem[{\citenamefont{Calmet et~al.}(2002)\citenamefont{Calmet, Jurco,
			Schupp, Wess, and Wohlgenannt}}]{Calmet:2001na}
	\bibinfo{author}{\bibfnamefont{X.}~\bibnamefont{Calmet}},
	\bibinfo{author}{\bibfnamefont{B.}~\bibnamefont{Jurco}},
	\bibinfo{author}{\bibfnamefont{P.}~\bibnamefont{Schupp}},
	\bibinfo{author}{\bibfnamefont{J.}~\bibnamefont{Wess}}, \bibnamefont{and}
	\bibinfo{author}{\bibfnamefont{M.}~\bibnamefont{Wohlgenannt}},
	\bibinfo{journal}{Eur. Phys. J.} \textbf{\bibinfo{volume}{C23}},
	\bibinfo{pages}{363} (\bibinfo{year}{2002}), \eprint{hep-ph/0111115}.
	
	\bibitem[{\citenamefont{Behr et~al.}(2003)\citenamefont{Behr, Deshpande,
			Duplancic, Schupp, Trampetic, and Wess}}]{Behr:2002wx}
	\bibinfo{author}{\bibfnamefont{W.}~\bibnamefont{Behr}},
	\bibinfo{author}{\bibfnamefont{N.~G.} \bibnamefont{Deshpande}},
	\bibinfo{author}{\bibfnamefont{G.}~\bibnamefont{Duplancic}},
	\bibinfo{author}{\bibfnamefont{P.}~\bibnamefont{Schupp}},
	\bibinfo{author}{\bibfnamefont{J.}~\bibnamefont{Trampetic}},
	\bibnamefont{and} \bibinfo{author}{\bibfnamefont{J.}~\bibnamefont{Wess}},
	\bibinfo{journal}{Eur. Phys. J.} \textbf{\bibinfo{volume}{C29}},
	\bibinfo{pages}{441} (\bibinfo{year}{2003}), \eprint{hep-ph/0202121}.
	
	\bibitem[{\citenamefont{Duplancic et~al.}(2003)\citenamefont{Duplancic, Schupp,
			and Trampetic}}]{Duplancic:2003hg}
	\bibinfo{author}{\bibfnamefont{G.}~\bibnamefont{Duplancic}},
	\bibinfo{author}{\bibfnamefont{P.}~\bibnamefont{Schupp}}, \bibnamefont{and}
	\bibinfo{author}{\bibfnamefont{J.}~\bibnamefont{Trampetic}},
	\bibinfo{journal}{Eur. Phys. J.} \textbf{\bibinfo{volume}{C32}},
	\bibinfo{pages}{141} (\bibinfo{year}{2003}), \eprint{hep-ph/0309138}.
	
	\bibitem[{\citenamefont{Hewett et~al.}(2001)\citenamefont{Hewett, Petriello,
			and Rizzo}}]{Hewett:2002zp}
	\bibinfo{author}{\bibfnamefont{J.~L.} \bibnamefont{Hewett}},
	\bibinfo{author}{\bibfnamefont{F.~J.} \bibnamefont{Petriello}},
	\bibnamefont{and} \bibinfo{author}{\bibfnamefont{T.~G.} \bibnamefont{Rizzo}},
	\bibinfo{journal}{eConf} \textbf{\bibinfo{volume}{C010630}},
	\bibinfo{pages}{E3064} (\bibinfo{year}{2001}), \eprint{hep-ph/0201275}.
	
	\bibitem[{\citenamefont{Mathews}(2001)}]{Mathews:2000we}
	\bibinfo{author}{\bibfnamefont{P.}~\bibnamefont{Mathews}},
	\bibinfo{journal}{Phys. Rev.} \textbf{\bibinfo{volume}{D63}},
	\bibinfo{pages}{075007} (\bibinfo{year}{2001}), \eprint{hep-ph/0011332}.
	
	\bibitem[{\citenamefont{Baek et~al.}(2001)\citenamefont{Baek, Ghosh, He, and
			Hwang}}]{Baek:2001ty}
	\bibinfo{author}{\bibfnamefont{S.}~\bibnamefont{Baek}},
	\bibinfo{author}{\bibfnamefont{D.~K.} \bibnamefont{Ghosh}},
	\bibinfo{author}{\bibfnamefont{X.-G.} \bibnamefont{He}}, \bibnamefont{and}
	\bibinfo{author}{\bibfnamefont{W.~Y.~P.} \bibnamefont{Hwang}},
	\bibinfo{journal}{Phys. Rev.} \textbf{\bibinfo{volume}{D64}},
	\bibinfo{pages}{056001} (\bibinfo{year}{2001}), \eprint{hep-ph/0103068}.
	
	\bibitem[{\citenamefont{Abbiendi et~al.}(2003)}]{Abbiendi:2003wva}
	\bibinfo{author}{\bibfnamefont{G.}~\bibnamefont{Abbiendi}} \bibnamefont{et~al.}
	(\bibinfo{collaboration}{OPAL}), \bibinfo{journal}{Phys. Lett.}
	\textbf{\bibinfo{volume}{B568}}, \bibinfo{pages}{181} (\bibinfo{year}{2003}),
	\eprint{hep-ex/0303035}.
	
	\bibitem[{\citenamefont{Yaser~Ayazi et~al.}(2012)\citenamefont{Yaser~Ayazi,
			Esmaeili, and Mohammadi-Najafabadi}}]{YaserAyazi:2012ni}
	\bibinfo{author}{\bibfnamefont{S.}~\bibnamefont{Yaser~Ayazi}},
	\bibinfo{author}{\bibfnamefont{S.}~\bibnamefont{Esmaeili}}, \bibnamefont{and}
	\bibinfo{author}{\bibfnamefont{M.}~\bibnamefont{Mohammadi-Najafabadi}},
	\bibinfo{journal}{Phys. Lett.} \textbf{\bibinfo{volume}{B712}},
	\bibinfo{pages}{93} (\bibinfo{year}{2012}), \eprint{1202.2505}.
	
	\bibitem[{\citenamefont{Alboteanu et~al.}(2006)\citenamefont{Alboteanu, Ohl,
			and Ruckl}}]{Alboteanu:2006hh}
	\bibinfo{author}{\bibfnamefont{A.}~\bibnamefont{Alboteanu}},
	\bibinfo{author}{\bibfnamefont{T.}~\bibnamefont{Ohl}}, \bibnamefont{and}
	\bibinfo{author}{\bibfnamefont{R.}~\bibnamefont{Ruckl}},
	\bibinfo{journal}{Phys. Rev.} \textbf{\bibinfo{volume}{D74}},
	\bibinfo{pages}{096004} (\bibinfo{year}{2006}), \eprint{hep-ph/0608155}.
	
	\bibitem[{\citenamefont{Ohl and Speckner}(2010)}]{Ohl:2010zf}
	\bibinfo{author}{\bibfnamefont{T.}~\bibnamefont{Ohl}} \bibnamefont{and}
	\bibinfo{author}{\bibfnamefont{C.}~\bibnamefont{Speckner}},
	\bibinfo{journal}{Phys. Rev.} \textbf{\bibinfo{volume}{D82}},
	\bibinfo{pages}{116011} (\bibinfo{year}{2010}), \eprint{1008.4710}.
	
	\bibitem[{\citenamefont{Selvaganapathy
			et~al.}(2016)\citenamefont{Selvaganapathy, Das, and
			Konar}}]{Selvaganapathy:2016jrl}
	\bibinfo{author}{\bibfnamefont{J.}~\bibnamefont{Selvaganapathy}},
	\bibinfo{author}{\bibfnamefont{P.~K.} \bibnamefont{Das}}, \bibnamefont{and}
	\bibinfo{author}{\bibfnamefont{P.}~\bibnamefont{Konar}},
	\bibinfo{journal}{Phys. Rev.} \textbf{\bibinfo{volume}{D93}},
	\bibinfo{pages}{116003} (\bibinfo{year}{2016}), \eprint{1602.02997}.
	
	\bibitem[{\citenamefont{Rafiei et~al.}(2017)\citenamefont{Rafiei, Rezaei, and
			Mirjalili}}]{Rafiei:2016lqq}
	\bibinfo{author}{\bibfnamefont{A.}~\bibnamefont{Rafiei}},
	\bibinfo{author}{\bibfnamefont{Z.}~\bibnamefont{Rezaei}}, \bibnamefont{and}
	\bibinfo{author}{\bibfnamefont{A.}~\bibnamefont{Mirjalili}},
	\bibinfo{journal}{Eur. Phys. J.} \textbf{\bibinfo{volume}{C77}},
	\bibinfo{pages}{319} (\bibinfo{year}{2017}), \eprint{1610.05259}.
	
	\bibitem[{\citenamefont{Chatrchyan et~al.}(2013)}]{Chatrchyan:2013tia}
	\bibinfo{author}{\bibfnamefont{S.}~\bibnamefont{Chatrchyan}}
	\bibnamefont{et~al.} (\bibinfo{collaboration}{CMS}), \bibinfo{journal}{JHEP}
	\textbf{\bibinfo{volume}{12}}, \bibinfo{pages}{030} (\bibinfo{year}{2013}),
	\eprint{1310.7291}.
	
	\bibitem[{\citenamefont{Khachatryan et~al.}(2015)}]{CMS:2014jea}
	\bibinfo{author}{\bibfnamefont{V.}~\bibnamefont{Khachatryan}}
	\bibnamefont{et~al.} (\bibinfo{collaboration}{CMS}), \bibinfo{journal}{Eur.
		Phys. J.} \textbf{\bibinfo{volume}{C75}}, \bibinfo{pages}{147}
	(\bibinfo{year}{2015}), \eprint{1412.1115}.
	
	\bibitem[{\citenamefont{Aaboud et~al.}(2017)}]{Aaboud:2017ffb}
	\bibinfo{author}{\bibfnamefont{M.}~\bibnamefont{Aaboud}} \bibnamefont{et~al.}
	(\bibinfo{collaboration}{ATLAS}), \bibinfo{journal}{JHEP}
	\textbf{\bibinfo{volume}{12}}, \bibinfo{pages}{059} (\bibinfo{year}{2017}),
	\eprint{1710.05167}.
	
	\bibitem[{\citenamefont{Aad et~al.}(2016)}]{Aad:2016zzw}
	\bibinfo{author}{\bibfnamefont{G.}~\bibnamefont{Aad}} \bibnamefont{et~al.}
	(\bibinfo{collaboration}{ATLAS}), \bibinfo{journal}{JHEP}
	\textbf{\bibinfo{volume}{08}}, \bibinfo{pages}{009} (\bibinfo{year}{2016}),
	\eprint{1606.01736}.
	
	\bibitem[{\citenamefont{Feldman and Cousins}(1998)}]{Feldman:1997qc}
	\bibinfo{author}{\bibfnamefont{G.~J.} \bibnamefont{Feldman}} \bibnamefont{and}
	\bibinfo{author}{\bibfnamefont{R.~D.} \bibnamefont{Cousins}},
	\bibinfo{journal}{Phys. Rev.} \textbf{\bibinfo{volume}{D57}},
	\bibinfo{pages}{3873} (\bibinfo{year}{1998}), \eprint{physics/9711021}.
	
\end{thebibliography}
\end{document}